%% file: main.tex
\documentclass[a4paper,11pt]{article}

\usepackage{caption}
\usepackage{subcaption}
\usepackage{pos}
\usepackage[super]{nth}
\usepackage{graphicx}
\usepackage{wrapfig,lipsum,booktabs}

\usepackage{lineno}
\title{Public Kaggle Competition ``IceCube -- Neutrinos in Deep Ice''}

\ShortTitle{IceCube Kaggle}

\author{The IceCube Collaboration \\{\normalsize \normalfont(a complete list of authors can be found at the end of the proceedings)}\\}

\emailAdd{philipp.eller@tum.de}

\abstract{

The reconstruction of neutrino events in the IceCube experiment is crucial for many scientific analyses, including searches for cosmic neutrino sources. The Kaggle competition ``IceCube -- Neutrinos in Deep ice'' was a public machine learning challenge designed to encourage the development of innovative solutions to improve the accuracy and efficiency of neutrino event reconstruction. Participants worked with a dataset of simulated neutrino events and were tasked with creating a suitable model to predict the direction vector of incoming neutrinos. From January to April 2023, hundreds of teams competed for a total of \$50k prize money, which was awarded to the best performing few out of the many thousand submissions. In this contribution I will present some insights into the organization of this large outreach project, and summarize some of the main findings, results and takeaways.
\vspace{4mm}
{\bfseries Corresponding author:}
Philipp Eller$^{1,2*}$\\
{$^{1}$ \itshape Technical University of Munich, TUM School of Natural Sciences, Physics Department, 85748 Garching, Germany}\\
{$^{2}$ \itshape Technical University of Munich, Munich Data Science Institute, Walther-von-Dyck-Strasse 10, 85748 Garching, Germany}\\[4mm]
$^*$ Presenter

\ConferenceLogo{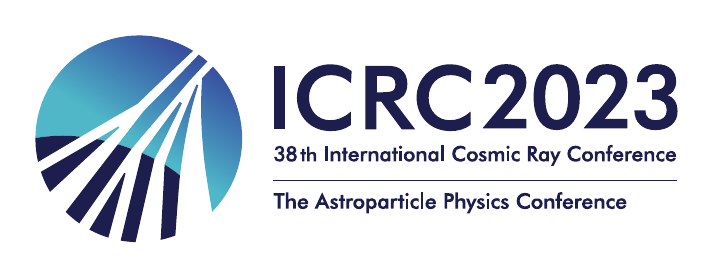}

\FullConference{The 38th International Cosmic Ray Conference (ICRC2023)\\ 26 July -- 3 August, 2023\\ Nagoya, Japan}
}

\begin{document}

\maketitle

\section{Introduction}\label{sec1}

``For a neutrino that interacted in IceCube, can you predict the direction it came from based on the detector readouts?'' is the question we asked thousands of people in the form of a public competition.
This task, described in further detail in the next section, is central to many physics analyses of IceCube~\cite{Aartsen:2016nxy} and similar detectors. The ability to estimate the neutrino's direction opens the door not only to search for neutrino point-sources \cite{IceCube:2018cha, IceCube:2022der}, but also to probe oscillation physics with atmospheric neutrinos \cite{IceCube:2023ewe,IceCube:2020phf}, to study the Earth absorption \cite{IceCube:2017roe}, or to search for dark matter \cite{IceCube:2021xzo} to name a few examples.
In fact, the question of this so-called ``event reconstruction'' is a long standing issue in the field and many brilliant minds have worked on it for decades.

In this proceedings contribution, we describe the organization and realization of a \href{https://www.kaggle.com}{kaggle}\footnote{https://www.kaggle.com} competition for IceCube. Section~\ref{sec:task} describes in greater detail the task participants were asked to solve, followed by Sec.~\ref{sec:orga} on the organization and implementation. Section~\ref{sec:live} talks about the phase when the competition was live, and the preliminary outcome is discussed in Sec.~\ref{sec:results}.

\section{Task: From Pulses to Directions}\label{sec1}
\label{sec:task}
\subsection{Dataset}

\begin{figure}[h]
    \centering
    \includegraphics[width=\textwidth]{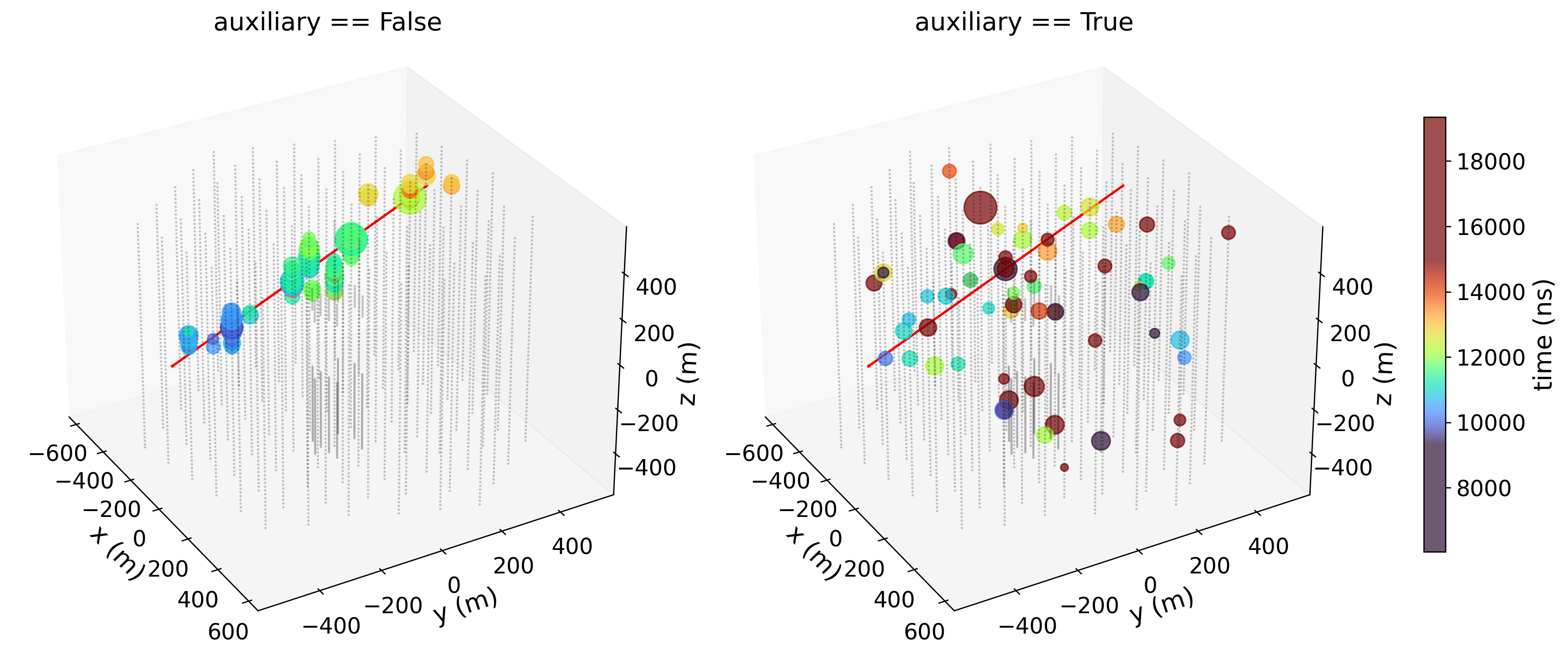}
    \caption{Example event from the kaggle datatset with a true azimuth of 4.86\,rad and zenith of 1.96\,rad.}
    \label{fig:example_event}
\end{figure}

The data from the IceCube neutrino observatory, after undergoing some basic processing, is in the form of a series of so-called ``pulses'' per triggered event~\cite{IceCube:2016zyt}. An example image visualizing these pulses for a particularly beautiful event is shown in Fig.~\ref{fig:example_event}. This series of pulses is a variable-length sequence, containing from few up to tens of thousand pulses per event, mainly depending on the interaction's energy. A pulse represents a number of inferred photo electrons in a photomultiplier tube (PMT) in the IceCube sensor array. Each pulse is characterized by the sensor ID that can be mapped to an $(x,y,z)$ coordinate, a time $t$, a charge $q$ that indicates the number of photo electrons represented by the pulse, and a flag here called ``auxiliary'' that identifies the readout mode. This readout mode is in more detail described in Ref.~\cite{IceCube:2016zyt} and in essence determines whether a pulse was in local coincidence to reduce PMT noise and read out by a fast waveform digitizer ($\textrm{aux}=\textrm{False}$), or not ($\textrm{aux}=\textrm{True}$).

What we are interested in in order to do science with such events, is to translate these pulse series into event-wise quantities that estimate properties such as the interacting neutrino's energy, the direction it came from, or its flavour. This step is known as event reconstruction. In our project we chose the neutrino's direction as the target variable for the reconstruction challenge to estimate.
To facilitate supervised machine learning approaches, and in order to be able to score solutions consistently, we based the competition on simulated neutrino events, i.e. synthetic data, that contains the knowledge of the true neutrino direction for each event that was input to the simulation. Figure~\ref{fig:example_event} also shows this true direction indicated by the red line.

The dataset provided to the participants, as well as the hidden dataset used for scoring the solutions, consist of 138 million simulated neutrino interactions of all flavours and energies between 100\,GeV up to 100\,PeV. The events were simulated following to a spectrum of $E^{-1.5}$ up to 1\,PeV and $E^{-1}$ for higher energies to increase the amount of the highest energy events.
Events were processed using the official IceCube tools, and required to pass simulated triggering and online filtering. 
But no further selection to reduce the sample to higher quality events was done. This means that the sample contains a large variety of events, ranging from noise-only pulse series, over lower energy events with few hits or events with coincident atmospheric muons, to spectacular high-energy tracks and cascades leaving stunning signatures in the detector.

\subsection{Scoring System}

The quality of a reconstructed direction can be expressed in terms of how far away it points from the true direction. This distance is geometrically the opening angle $\Psi$ between the true direction $(\varphi_\textrm{true}, \vartheta_\textrm{true})$ and reconstructed direction $(\varphi_\textrm{reco}, \vartheta_\textrm{reco})$ expressed in spherical coordinates azimuth ($\varphi$) and zenith ($\vartheta$), and given below:
\newcommand{\sz}[1]{\sin{\vartheta_\textrm{#1}}}
\newcommand{\cz}[1]{\cos{\vartheta_\textrm{#1}}}
\newcommand{\sa}[1]{\sin{\varphi_\textrm{#1}}}
\newcommand{\ca}[1]{\cos{\varphi_\textrm{#1}}}
\begin{equation}
    \Psi = \arccos \left( \sz{true} \sz{reco} (\ca{true}\ca{reco} + \sa{true}\sa{reco}) + \cz{true}\cz{reco} \right).
\end{equation}
To turn this event-wise quantity into an overall score, we compute the simple mean of it over an entire dataset---hence forth referred to as the the mean angular error. The smaller this number, the better the overall reconstruction quality. The mean angular error computed over the hidden scoring set represents the competition's ranking system, and the team with the smallest score achieved wins.

Out of the 138 million events in the dataset, 137 million were provided to the participants to download including the truth information. The remaining one million events scoring set was retained for computing the ranks. This scoring happened in an isolated system without connection to the outside. Participants had to upload their reconstruction algorithm as a jupyter notebook (including any additional files needed, e.g. pre-trained neural network weights), which then was run on the separate system on the scoring set.
Execution was limited to a total of nine hours compute time, which in turn means a minimum inference speed of 31\,Hz including I/O. In comparison, the total trigger rate of IceCube is $2.7\pm0.2$\,kHz \cite{IceCube:2016zyt}, which means that any valid solution will be able to process the entire IceCube online data stream with at most 100 compute instances. 

The scoring set was further subdivided into two roughly equal parts. On one part the ``public'' score was computed, that was visible to anyone during the competition on the public leader board (LB). The other half remained sealed off until the competition finished, and the scores computed on this constitute the ``private'' LB that determined the prize winners. This extra separation is in place to prevent overfitting on the public LB score.

\section{Organization}
\label{sec:orga}

The organization and realization of this IceCube project was carried out in close collaboration with several institutions. The preparation, realization and hosting of this project was significantly supported by kaggle, including the help of a project manager, a data scientist and the provision of their platform infrastructure and compute resources, as well as dissemination channels.
At the same time the project was supported by several science institution: The \href{https://www.tum.de}{Technical University Munich}\footnote{https://www.tum.de}, its \href{https://www.mdsi.tum.de}{Munich Data Science Institute}\footnote{https://www.mdsi.tum.de}, the 
\href{https://www.sfb1258.de}{Collaborative Research Center SFB 1258}\footnote{https://www.sfb1258.de}, the \href{https://www.origins-cluster.de}{Excellence Cluster ORIGINS}\footnote{https://www.origins-cluster.de}, and the \href{https://www.punch4nfdi.de}{PUNCH4NFDI}\footnote{https://www.punch4nfdi.de} consortium all helped in the preparatory phase, the project realization, and dissemination.
The hosting by kaggle included the \href{https://www.kaggle.com/competitions/icecube-neutrinos-in-deep-ice}{competition page}\footnote{https://www.kaggle.com/competitions/icecube-neutrinos-in-deep-ice} \cite{icecube-neutrinos-in-deep-ice} that contains the project description, the dataset, leader board, discussion forum, code collection and further information, and remains available as a learning resource.

In order to attract a large number of participants to guarantee a success of the project, the following incentives in the form of cash prizes were created:
\begin{itemize}\setlength\itemsep{0.1em}
    \item \textbf{Early Sharing Prize}: \$\,5,000 for the best solution two weeks into the competition that was shared publicly, i.e. free for other participants to use.
    \item \textbf{Leader board Top 3}: \$\,18,000, \$\,12,000, and \$\,10,000 awarded to the \nth{1}, \nth{2}, and \nth{3} place, respectively, at the end of the competition, given the private leader board score
    \item \textbf{Solution write-ups}: 5$\times$ \$\,1,000 for the five most interesting solution write-ups among the top 30, two weeks after the competition close.
\end{itemize}

\section{Competition Phase}
\label{sec:live}

\begin{figure}[h]
    \centering
    \subfloat[][Participants per country (image courtesy of kaggle)]{\includegraphics[width=0.65\textwidth]{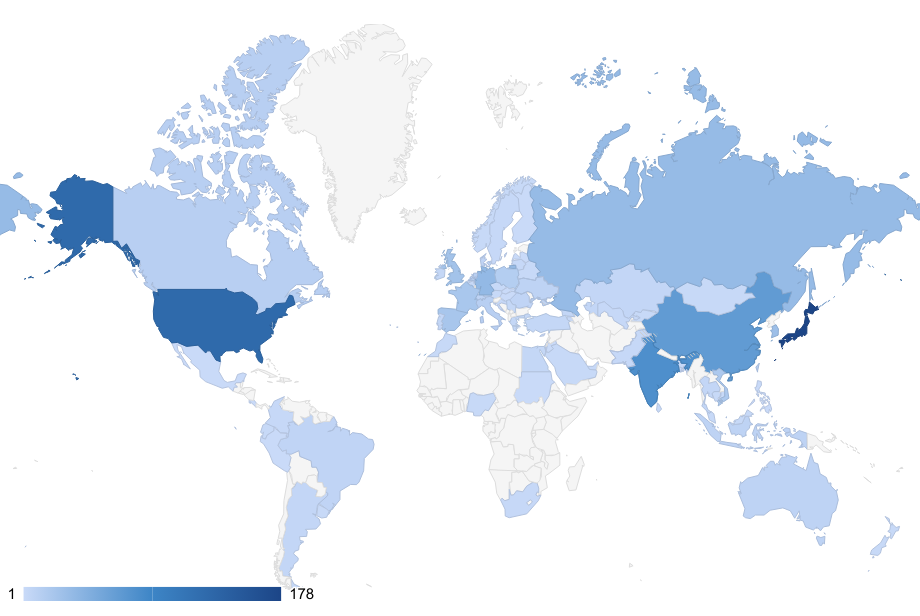}
    }
    %\qquad
    \subfloat[][Top 5 countries participant numbers and percentage of total]{
    \small
\begin{tabular}{lll}
\multicolumn{3}{l}{\textbf{Top 5 Countries}}\\\hline
Japan & 182 & 19.89\% \\
USA & 133 & 14.54\% \\
India & 81 & 8.85\% \\
China & 62 & 6.78\% \\
Germany & 34 & 3.72\% \\
\end{tabular}
}    
    \caption{Distribution of the 901 participants per country of residence.}
    \label{fig:worldmap}
\end{figure}

The competition went live on January 19, 2023 and ran for 3 months until April 19, 2023.
During the course of these three months, a total of 6,460 people registered and signed up for our IceCube competition. 
Among these registered users, a total of 901 participants handed in at least one valid solution, 100 of which were first time kaggle users. At competition close, participants had organized themselves in 812 teams. 
Over the entire course of the three months, a grand total of 11,206 solutions had been submitted.
The project attracted competitors around the globe from 74 countries. Figure~\ref{fig:worldmap}a show the distribution of participants on a political world map, and Table~\ref{fig:worldmap}b summarizes the top 5 countries in terms of total number of participants.

\begin{figure}[h]
    \centering
    \includegraphics[width=\textwidth]{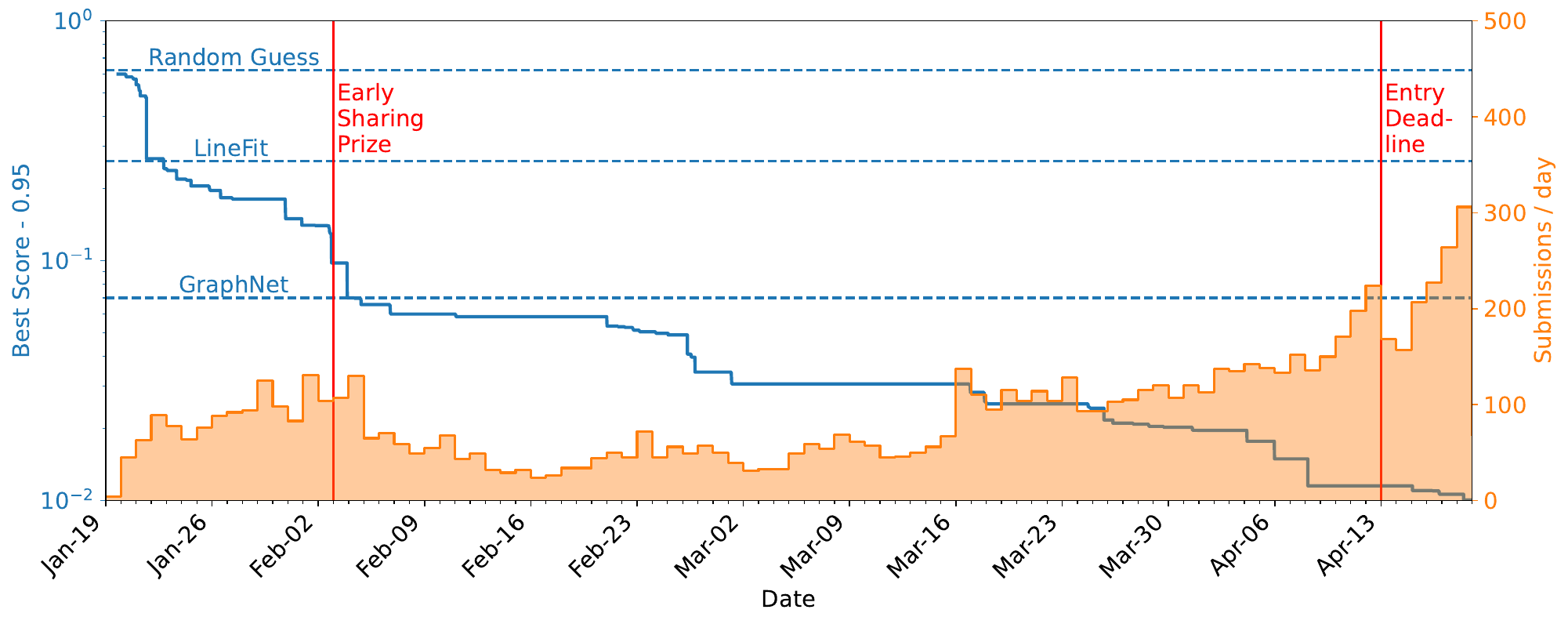}
    \caption{Development of the best score as a function of the time during the competition (blue color, left y-axis) and number of submissions per day (orange color, right y-axis).}
    \label{fig:timeline}
\end{figure}

Figure~\ref{fig:timeline} provides some metrics as a function of time. The evolution of the score (mean angular error, lower = better) is starting at a value of around $\pi/2$ which corresponds to the score of a random guess. After a few days into the competition, solutions that are already comparable to a simple online reconstruction, used by IceCube, called \textsc{LineFit} \cite{linefit} were achieved. The Early Sharing Prize (ESP) winning solution two weeks into the competition was based on the open source \textsc{GraphNeT} repository \cite{Søgaard2023} that had initially been developed to reconstruct low-energy (< 100\,GeV) IceCube neutrinos \cite{IceCube:2022njh}. An official example notebook for \textsc{GraphNeT} was released shortly after the ESP.
The score developed further, steadily getting better until the competition end. Figure~\ref{fig:timeline} also shows the amount of submissions that were scored per day, exhibiting some elevated numbers around the ESP, and then especially towards the competition end peaking at over 300 submissions per day.

\section{Outcome, Conclusions \& Outlook}
\label{sec:results}

While the defined goal of the competition was to find a novel reconstruction method, of which the outcome will be discussed below, there have been other outputs. The \href{https://www.kaggle.com/competitions/icecube-neutrinos-in-deep-ice/discussion}{discussion forum}\footnote{https://www.kaggle.com/competitions/icecube-neutrinos-in-deep-ice/discussion} of our competition page was extensively used by the community and also the hosts. 
\begin{wrapfigure}{r}{0.6\textwidth}
    \includegraphics[width=0.6\textwidth]{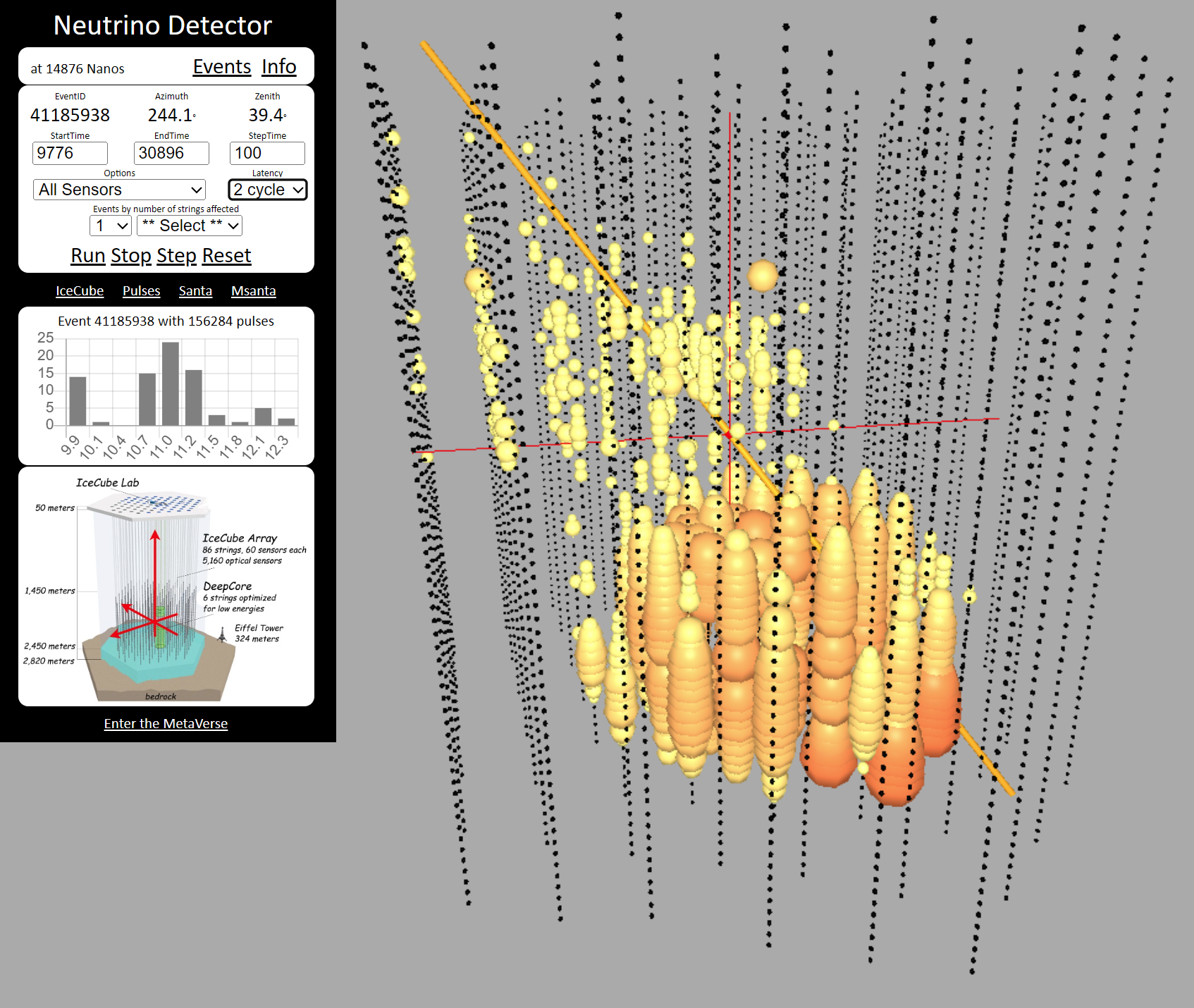}
    \caption{Interactive, 3-dimensional, animated event viewer, available \href{https://crumbsoftware.com:3028/meta.htm?object=icecube}{here}. (Courtesy of Ed Unverricht)}
    \label{fig:event-viewer}
\end{wrapfigure}
A total of 177 separate discussion threads were started, with 651 discussion comments. These discussions concerned organizational topics, neutrino physics, IceCube related questions and discussions, machine learning, and more.
On the competitions \href{https://www.kaggle.com/competitions/icecube-neutrinos-in-deep-ice/code}{code page}\footnote{https://www.kaggle.com/competitions/icecube-neutrinos-in-deep-ice/code} 322 jupyter notebooks were published by kaggle community members. These notebooks contain example code snippets, exploratory data analyses, event viewers, data loading and pre-processing steps, and so forth.
Figure~\ref{fig:event-viewer} shows a screenshot of a particularly impressive, interactive 3d event viewer created by a user that runs in the web browser and allows to inspect events in the kaggle IceCube dataset.

\begin{figure}[h]
     \centering
     \begin{subfigure}[b]{0.49\textwidth}
         \centering
         \includegraphics[width=\textwidth]{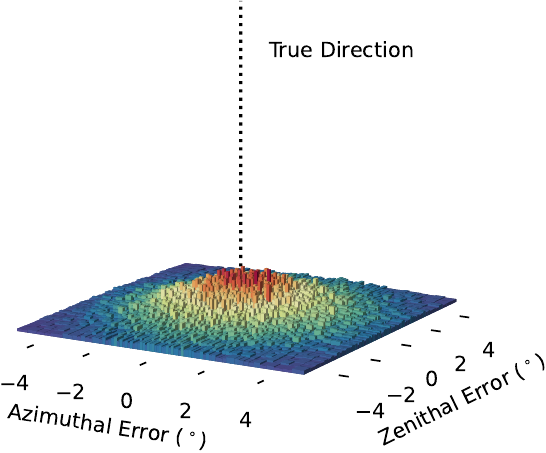}
         \caption{Early Sharing Prize (ESP)}
         \label{fig:lego_esp}
     \end{subfigure}
     \hfill
     \begin{subfigure}[b]{0.49\textwidth}
         \centering
         \includegraphics[width=\textwidth]{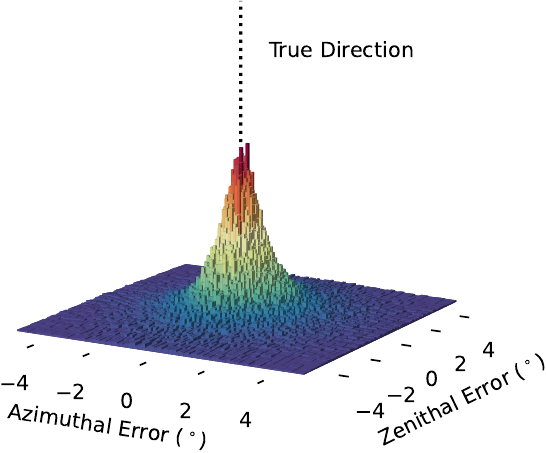}
         \caption{\nth{1} Place}
         \label{fig:lego_1st}
     \end{subfigure}\\[3ex]
     \begin{subfigure}[b]{0.49\textwidth}
         \centering
         \includegraphics[width=\textwidth]{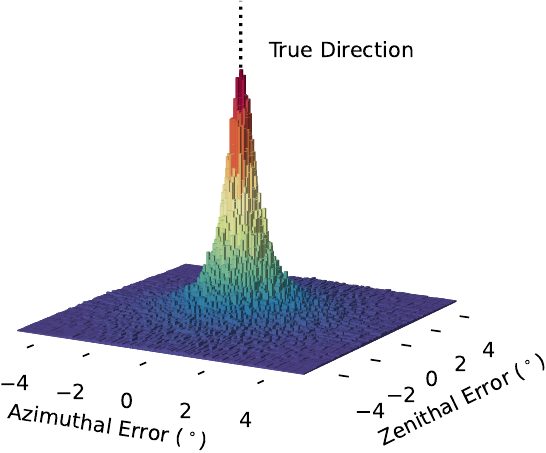}
         \caption{\nth{2} Place}
         \label{fig:lego_2nd}
     \end{subfigure}
     \hfill
     \begin{subfigure}[b]{0.49\textwidth}
         \centering
         \includegraphics[width=\textwidth]{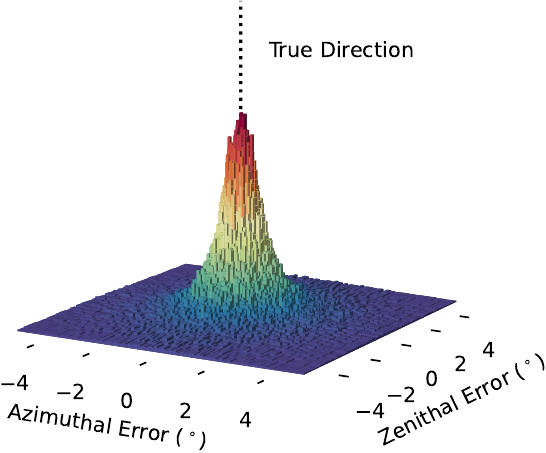}
         \caption{\nth{3} Place}
         \label{fig:lego_3rd}
     \end{subfigure}
        \caption{Distribution of the angular error over the entire scoring set in terms of zenith and azimuth error. Shown is a zoom in to the $\pm5^\circ$ region around the true direction only.}
        \label{fig:lego}
\end{figure}

\begin{wrapfigure}{r}{0.5\textwidth}
    \centering
    \includegraphics[width=0.5\textwidth]{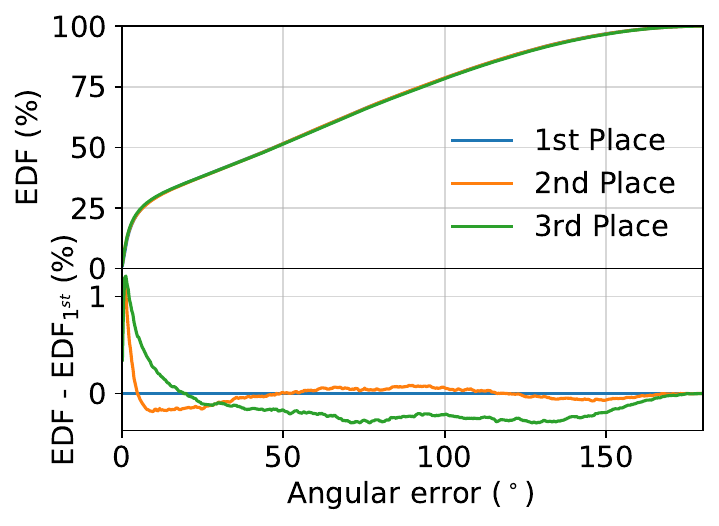}
    \caption{Empirical cumulative distributions of the top 3 solutions on the scoring dataset in terms of angular error. The top panel shows the EDFs , while the bottom one shows the difference to the \nth{1} place.}
    \label{fig:edf}
\end{wrapfigure}

The outcome of the main challenge is a set of machine learning based reconstruction methods that achieve a sub-degree track resolution for the IceCube dataset and are applicable to any event regardless of its signature. In Fig.~\ref{fig:lego} the distribution of events within $\pm5^\circ$ of the true direction are shown for the top 3 solutions and the ESP, i.e. all prize money winners.
A stark difference and evolution from the ESP compared to the top 3 can be observed, highlighting the improvements that were made by the competitors over the course of the competition. The figure also shows a slight difference in performance among the top 3 solutions. One thing to note is that the \nth{2} place on the private LB is corresponding to the \nth{1} place on the public LB and vice versa, illustrating how close these scores are. Furthermore, Fig.~\ref{fig:lego} only shows the performance in the region $\pm5^\circ$, while many events reconstruct to larger angels, especially cascades. In Fig.~\ref{fig:edf} a more complete picture over the full scoring data set is provided, showing the empirical cumulative distribution (EDF) of the angular error for the top 3. It can be seen that the \nth{2} place reconstructs more events within an angle of around $5^\circ$ but then is outperformed by the \nth{1} place. For the \nth{3} place this transition happens at around $20^\circ$.
This means that the top 3 solutions differ in terms of what events they reconstruct best, and that leveraging a combination of the methods may further improve overall performance.

The methods behind the winning solutions -- including those of several top-20 solutions --  are documented in write-ups available on the \href{https://www.kaggle.com/competitions/icecube-neutrinos-in-deep-ice/leaderboard}{leader board}\footnote{https://www.kaggle.com/competitions/icecube-neutrinos-in-deep-ice/leaderboard}. The teams of the \nth{1}, \nth{2}, \nth{3}, \nth{5}, and \nth{11} were selected for the additional \$1,000 prize money each for the most interesting solution write-ups.

What the top 3 solutions and many other top performing solutions share in common, is that their models include, at least in parts, a transformer architecture implementing an attention mechanism \cite{vaswani2017attention}. This constitutes an entirely new approach to process IceCube data, and demonstrably works well outperforming many other architectures (linear models, BDTs, CNNs, GNNs, RNNs, etc.) used by kaggle competitors, but also IceCube scientists.

In summary, the public kaggle competition ``IceCube -- Neutrinos in Deep Ice'' ran from January 19 until April 19, 2023, attracting thousands of people from around the world. More than 800 teams in the end competed for the total of \$50,000 prize money. During the competition, countless discussions happened and data visualizations and example codes were generated and shared, sparking an interest in neutrino physics and IceCube.
From an outreach point of view, the competition was therefore a big success and the competition page remains available as a great learning resource.
From the scientific point of view, new solutions for the reconstruction of neutrino directions were found, many based on the popular transformer architecture.
The top solutions are able to localize well-reconstructable events well within a degree of error, and can do that at inference speeds of the order of $\mathcal{O}(100\,\textrm{Hz})$. This opens the possibility to employ these methods not only for offline data analyses, but even for online event processing.

The next steps will encompass an implementation of the winning solutions in the IceCube software stack, eventually a combination of the top solutions into an even more capable one, and detailed comparisons to existing IceCube approaches. In the longer term, the hope is that these approaches can be used to improve IceCube processing and ultimately science results.

% Bibtex references:
\bibliographystyle{ICRC}
\bibliography{references}

% Alternatively, you can include references by hand:
%\begin{thebibliography}{99}
%\bibitem{...}
%
%\end{thebibliography}

\clearpage

%The following list of authors, affiliations and funding agencies will be updated at the day of submission. The following template is a placeholder generated via https://authorlist.icecube.wisc.edu/icecube on March 25, 2023 and will be updated.
\input{authorlist_IceCube.tex}

\end{document}

%% file: authorlist_IceCube.tex
\section*{Full Author List: IceCube Collaboration}

\scriptsize
\noindent
R. Abbasi$^{17}$,
M. Ackermann$^{63}$,
J. Adams$^{18}$,
S. K. Agarwalla$^{40,\: 64}$,
J. A. Aguilar$^{12}$,
M. Ahlers$^{22}$,
J.M. Alameddine$^{23}$,
N. M. Amin$^{44}$,
K. Andeen$^{42}$,
G. Anton$^{26}$
% ...
\\
\\
% ...
$^{12}$ Universit{\'e} Libre de Bruxelles, Science Faculty CP230, B-1050 Brussels, Belgium \\
% ...
$^{17}$ Department of Physics, Loyola University Chicago, Chicago, IL 60660, USA \\
$^{18}$ Dept. of Physics and Astronomy, University of Canterbury, Private Bag 4800, Christchurch, New Zealand \\
% ...
$^{22}$ Niels Bohr Institute, University of Copenhagen, DK-2100 Copenhagen, Denmark \\
$^{23}$ Dept. of Physics, TU Dortmund University, D-44221 Dortmund, Germany \\
% ...
$^{26}$ Erlangen Centre for Astroparticle Physics, Friedrich-Alexander-Universit{\"a}t Erlangen-N{\"u}rnberg, D-91058 Erlangen, Germany \\
% ...
$^{40}$ Dept. of Physics and Wisconsin IceCube Particle Astrophysics Center, University of Wisconsin{\textendash}Madison, Madison, WI 53706, USA \\
% ...
$^{42}$ Department of Physics, Marquette University, Milwaukee, WI, 53201, USA \\
% ...
$^{44}$ Bartol Research Institute and Dept. of Physics and Astronomy, University of Delaware, Newark, DE 19716, USA \\
% ...
$^{63}$ Deutsches Elektronen-Synchrotron DESY, Platanenallee 6, 15738 Zeuthen, Germany  \\
$^{64}$ Institute of Physics, Sachivalaya Marg, Sainik School Post, Bhubaneswar 751005, India \\
% ...

\subsection*{Acknowledgements}

\noindent
The authors gratefully acknowledge the support from the following agencies and institutions:
USA {\textendash} U.S. National Science Foundation-Office of Polar Programs,
U.S. National Science Foundation-Physics Division,
U.S. National Science Foundation-EPSCoR,
Wisconsin Alumni Research Foundation,
Center for High Throughput Computing (CHTC) at the University of Wisconsin{\textendash}Madison,
Open Science Grid (OSG),
Advanced Cyberinfrastructure Coordination Ecosystem: Services {\&} Support (ACCESS),
Frontera computing project at the Texas Advanced Computing Center,
U.S. Department of Energy-National Energy Research Scientific Computing Center,
Particle astrophysics research computing center at the University of Maryland,
Institute for Cyber-Enabled Research at Michigan State University,
and Astroparticle physics computational facility at Marquette University;
Belgium {\textendash} Funds for Scientific Research (FRS-FNRS and FWO),
FWO Odysseus and Big Science programmes,
and Belgian Federal Science Policy Office (Belspo);
Germany {\textendash} Bundesministerium f{\"u}r Bildung und Forschung (BMBF),
Deutsche Forschungsgemeinschaft (DFG),
Helmholtz Alliance for Astroparticle Physics (HAP),
Initiative and Networking Fund of the Helmholtz Association,
Deutsches Elektronen Synchrotron (DESY),
and High Performance Computing cluster of the RWTH Aachen;
Sweden {\textendash} Swedish Research Council,
Swedish Polar Research Secretariat,
Swedish National Infrastructure for Computing (SNIC),
and Knut and Alice Wallenberg Foundation;
European Union {\textendash} EGI Advanced Computing for research;
Australia {\textendash} Australian Research Council;
Canada {\textendash} Natural Sciences and Engineering Research Council of Canada,
Calcul Qu{\'e}bec, Compute Ontario, Canada Foundation for Innovation, WestGrid, and Compute Canada;
Denmark {\textendash} Villum Fonden, Carlsberg Foundation, and European Commission;
New Zealand {\textendash} Marsden Fund;
Japan {\textendash} Japan Society for Promotion of Science (JSPS)
and Institute for Global Prominent Research (IGPR) of Chiba University;
Korea {\textendash} National Research Foundation of Korea (NRF);
Switzerland {\textendash} Swiss National Science Foundation (SNSF);
United Kingdom {\textendash} Department of Physics, University of Oxford.

%% file: main.bbl
\providecommand{\href}[2]{#2}\begingroup\raggedright\begin{thebibliography}{10}

\bibitem{Aartsen:2016nxy}
{\bfseries IceCube} Collaboration, M.~G. Aartsen {\em et~al.}
  \href{http://dx.doi.org/10.1088/1748-0221/12/03/P03012}{{\em JINST}
  {\bfseries 12} no.~03, (2017) P03012}.

\bibitem{IceCube:2018cha}
{\bfseries IceCube} Collaboration, M.~G. Aartsen {\em et~al.}
  \href{http://dx.doi.org/10.1126/science.aat2890}{{\em Science} {\bfseries
  361} no.~6398, (2018) 147--151}.

\bibitem{IceCube:2022der}
{\bfseries IceCube} Collaboration, R.~Abbasi {\em et~al.}
  \href{http://dx.doi.org/10.1126/science.abg3395}{{\em Science} {\bfseries
  378} no.~6619, (2022) 538--543}.

\bibitem{IceCube:2023ewe}
{\bfseries IceCube} Collaboration, R.~Abbasi {\em et~al.}
  \href{https://arxiv.org/abs/2304.12236}{arXiv:2304.12236}.

\bibitem{IceCube:2020phf}
{\bfseries IceCube} Collaboration, M.~G. Aartsen {\em et~al.}
  \href{http://dx.doi.org/10.1103/PhysRevLett.125.141801}{{\em Phys. Rev.
  Lett.} {\bfseries 125} no.~14, (2020) 141801}.

\bibitem{IceCube:2017roe}
{\bfseries IceCube} Collaboration, M.~G. Aartsen {\em et~al.}
  \href{http://dx.doi.org/10.1038/nature24459}{{\em Nature} {\bfseries 551}
  (2017) 596--600}.

\bibitem{IceCube:2021xzo}
{\bfseries IceCube} Collaboration, R.~Abbasi {\em et~al.}
  \href{http://dx.doi.org/10.1103/PhysRevD.105.062004}{{\em Phys. Rev. D}
  {\bfseries 105} no.~6, (2022) 062004}.

\bibitem{IceCube:2016zyt}
{\bfseries IceCube} Collaboration, M.~G. Aartsen {\em et~al.}
  \href{http://dx.doi.org/10.1088/1748-0221/12/03/P03012}{{\em JINST}
  {\bfseries 12} no.~03, (2017) P03012}.

\bibitem{icecube-neutrinos-in-deep-ice}
A.~Chow, L.~Heinrich, P.~Eller, R.~{\O}rs{\o}e, and S.~Dane, ``{IceCube -
  Neutrinos in Deep Ice},'' 2023.
\newblock \url{https://kaggle.com/competitions/icecube-neutrinos-in-deep-ice}.

\bibitem{linefit}
{\bfseries IceCube} Collaboration.
  \href{https://inspirehep.net/files/b0732cd83ca95f49be68b24e8be7e0c3}{33rd
  International Cosmic Ray Conference (ICRC2013)}.

\bibitem{Søgaard2023}
A.~Søgaard {\em et~al.} \href{http://dx.doi.org/10.21105/joss.04971}{{\em
  Journal of Open Source Software} {\bfseries 8} no.~85, (2023) 4971}.

\bibitem{IceCube:2022njh}
{\bfseries IceCube} Collaboration, R.~Abbasi {\em et~al.}
  \href{http://dx.doi.org/10.1088/1748-0221/17/11/P11003}{{\em JINST}
  {\bfseries 17} no.~11, (2022) P11003}.

\bibitem{vaswani2017attention}
A.~Vaswani {\em et~al.}
  \href{https://arxiv.org/abs/1706.03762}{arXiv:1706.03762}.

\end{thebibliography}\endgroup
